\documentclass[%
 reprint,
 amsmath,amssymb,
 aps,
]{revtex4-1}
 \usepackage{color}
\usepackage{graphicx}
\usepackage{dcolumn}
\usepackage{bm}

\begin{document}

\preprint{APS/123-QED}

\title{Finite- Size Scaling of Correlation Function}

\author{Xin Zhang$^{1,3}$}{}
\author{Gaoke Hu$^{1,3}$}{}
\author{Yongwen Zhang$^{1,2}$}
\author{Xiaoteng Li$^{1,3}$}
\author{Xiaosong Chen$^{1,3}$}%
\email{Email address: chenxs@itp.ac.cn }
\address{$^1$Institute of Theoretical Physics, Key Laboratory of Theoretical Physics, Chinese Academy of Science, P.O. Box 2735, Beijing, 100190, China\\
$^2$Data Science Research Center, Kunming University of Science and Technology, Kunming, 100190, Yunnan,  China\\
$^3$School of Physical Sciences, University of Chinese Academy of Science,  No. 19A Yuquan Road, Beijing 100049, China
}

\date{\today}

\begin{abstract}
We propose the finite-size scaling of correlation function in a finite system near its critical point. At a distance ${\bf r}$ in the finite system with size $L$, the correlation function can be written as the product of $|{\bf r}|^{-(d-2+\eta)}$ and its finite-size scaling function of variables ${\bf r}/L$ and $tL^{1/\nu}$, where $t=(T-T_c)/T_c$. The directional dependence of correlation function is nonnegligible only when $|{\bf r}|$ becomes compariable with $L$.  This finite-size scaling of correlation function has been confirmed by correlation functions of the Ising model and the bond percolation in two-diemnional lattices, which are calculated by Monte Carlo simulation. We can use the finite-size scaling  of correlation function to determine the critical point and the critical exponent $\eta$.
\begin{description}
\item[PACS numbers]
64.60.-i, 05.50.+q
\item[keywords]
critical phenomena, finite-size scaling, correlation function, lattice model
\end{description}
\end{abstract}

\pacs{05.70.Jk} 
\keywords{Suggested keywords}
\maketitle


\section{ Introduction}

The concept of finite-size scaling has played an important role in the investigation of finite-size effects near critical point over last decades\cite{Privman1991}. The free-energy dendity $f(t,L)$ at the reduced temperature $t=(T - T_c)/T_c$ and vanisheing external field in a $d$-dimensional cubic geometry of volume $L^d$ with periodic boundary conditions (PBC) can be decomposed as
\begin{equation}
\label{free-energy}
f(t,L) = f_s (t,L) + f_0 (t)\;,
\end{equation}
where $f_s (t,L)$ denotes the singular part of $f$ and $f_0 (t)$ is the regular part. It was asserted by Provman and Fisher \cite{PhysRevB.30.322} that, below the upper critical dimension $d=4$, $f_s (t,L)$ has the asymptotic finite-size scaling structure
\begin{equation}
\label{fscaling}
f_s (t,L) = L^{-d} F (tL^{1/\nu})\;,
\end{equation}
where $F(x)$ ia a universal scaling function. 

For the free energy density in the bulk limit $L \to \infty$,  it was believed that there is the two-scale-factor universality in general \cite{Privman1991}. It has been demonstrated that the two-scale-factor universality is valid for isotropic systems but not for anisotropic systems with noncubic symmetry \cite{ISI:000225970700048}. The finite-size scaling function of the free energy density $F$ depends on the normalized anisotropic matrix $\bar{\bf A}$ of anisotropic systems \cite{ISI:000225970700048}. The Binder cumulant ratio $U$ was introduced \cite{PhysRevLett.47.693} to determine the transition point and compute the critical exponent of the correlation length. Its value at the transition point $U(T_c)$ was believed to be universal and can indicate the the universality class together with critical exponents \cite{Privman1991}. It has been proved in Ref.\cite{ISI:000225970700048} that $U(T_c)$ depends on $\bar{\bf A}$ also.  Using the Monte Carlo simulations of two-dimensional anisotropic Ising models, the dependence of $U(T_c)$ on $\bar{\bf A}$ has been confirmed \cite{0305-4470-38-44-L03,PhysRevE.80.042104}. The experimental investigations of the finite-size scaling have been done extensively with $^4He$ at the superfluid transition and are reviewed in Refs.\cite{RevModPhys.79.1,RevModPhys.80.1009}.

A similar finite-size scaling ansatz of the correlation length $\xi (t,L)$ was made by Privman and Fisher \cite{PhysRevB.30.322} as
\begin{equation}
\label{xi-scaling}
\xi (t,L) = L X(tL^{1/\nu})\;,
\end{equation}
with the universal scaling function $X(x)$. In the three-dimensional Ising model, the finite-size scaling analysis of the correlation length with Monte Carlo simulations was reported \cite{PhysRevLett.62.2608}. Above the upper critcal dimension, the finite-size scaling of the correlation length was studied in the five-dimensional Ising model \cite{PhysRevB.71.174438}.  As the central quantity in a statistical system, the correlaion length in ionic fluids was investigated by integral equation theory \cite{Guoyy2010}. The finite-size scaling can be related to the large distance behavior of bulk order-parameter correlaion function  \cite{Chen2000}.

In this paper, we study the correlation function $g({\bf r}, t,L)$ of finite system with size $L$. In Sec. II, the finite-size scaling structure of correlation function $g({\bf r},t,L)$ is introduced and discussed. With the correlation function of Ising model in Sec. III and the correlation function of bond percolation In Sec. IV, the finite-size scaling of correlation function is confirmed by Monte Carlo simulation. In Sec. V, we draw some conclusions.

\section{Finite-size scaling of correlation function }

Correlation functions describe how microscopic variables at different positions are related. In an infinite system with a lattice spacing $\tilde{a}$ , the correlation function $g ({\bf r},t,\infty)$ reads \cite{Privman1991,PhysRevB.30.322} for $|t|\ll 1$ and $|{\bf r}| \gg \tilde{a}$,
\begin{equation}
g({\bf r},t,\infty) = A\; |{\bf r}|^{-d+2-\eta}\Phi (|{\bf r}|/\xi)
\end{equation}  
with a universal scaling function $\Phi$, a nonuniversal amplitude $A$, and with $\xi =\xi_0 |t|^{-\nu}$, apart from correlations to scaling. 

In a finite system with $N$ spins, the correlation $C_{ij}$ between spin $i$ positioned at ${\bf r}_i$ and spin $j$ positioned at ${\bf r}_j$ can be calculated as \cite{0253-6102-66-3-355} 
\begin{eqnarray}
\label{cor}
C_{ij} = \langle S_i  S_j\rangle - \langle S_i \rangle \langle S_j \rangle \;,
\end{eqnarray}
where the averages are done over all configurations with weight  $p(\{S_i\})$.
Using $C_{ij}$ as its elements, a $N \times N$ correlation matrix ${\bf C}$ can be obtained. $C$ has $N$ eigenvectors and eigenvalues. For eigenvector ${\bf b}_n$ of eigenvalue $\lambda_n$, there is the relation
\begin{equation}
\label{eigen}
{\bf C} {\bf b}_n = \lambda_n {\bf b}_n, \; n=1,2,...,N,
\end{equation}
where
\begin{equation}
{\bf b}_n = \left[ \begin{array}{c} b_{1n} \\ b_{2n} \\.\\.\\.\\b_{Nn}\end{array} \right]\;.
\end{equation}
The normalized eigenvectors are orthogonal to each other and satisfy the conditions
\begin{equation}
\label{ortho}
{\bf b}_n \cdot {\bf b}_l =\sum_{j} b_{jn} b_{jl}=\delta_{nl}\;,
\end{equation}
where $\delta_{nl}$ is the Kronecker delta.

From the fluctuations of $N$ spins, we can define $N$ principal fluctuation modes
\begin{equation}
\label{mode}
\delta \widetilde{S}_n = \sum_{j=1}^{N} b_{jn} \delta S_j, \;n=1,2,...,N\;.
\end{equation}
Using the orthogonal conditions in Eq.\ref{ortho}, the correlation between principal fluctuation modes can be calculated as
\begin{equation}
\label{ortho_lambda}
\widetilde{C}_{nl} \equiv \left<  \delta \widetilde{S}_n \; \delta \widetilde{S}_l \right> = \lambda_n \delta_{nl}\;.
\end{equation}
There is no correlation between different principal fluctuation modes. The mean square of principal fluctuation mode $\delta \widetilde{S}_n$ is equal to $\lambda_n$.

We can express the correlation $C_{ij}$ between spin $i$ and spin $j$ by eigenvectors and eigenvalues as
\begin{equation}
C_{ij} = \sum_{n=1}^{N} b_{in} b_{jn} \lambda_n\;.
\end{equation}
We introduce $C_{ij}^{(n)}= b_{in}b_{jn}$, which can be understood as the correlation between spin $i$ and spin $j$ in the $n$-th principal fluctuation mode. The total correlation between spin $i$ and spin $j$ is obtained by summing correlations of all principal fluctuation modes as
\begin{equation}
C_{ij}=\sum_{n=1}^{N} \lambda_n C_{ij}^{(n)}\;.
\end{equation}

In Ref.\cite{0253-6102-66-3-355}, it has been proposed and confirmed by the two-diemnsional Ising model that eigenvalues satisfy the following finite-size scaling form
\begin{equation}
\label{lambdas}
\lambda_n (t,L) = L^{2-\eta} f_n (t L^{1/\nu})\;,
\end{equation} 
where $\eta$ is related to the critical exponent $\gamma$ of susceptibility by the hypescaling relation $2-\eta=\gamma/\nu$. 
Using position vectors ${\bf r}_i$ of spins, eigenvector ${\bf b}_n$ can represented by a spatial function $b_n ({\bf r}_i)$.  In a finite system of volume $L^d$ and periodic boundary conditions, the spatial function can be expressed as
\begin{equation}
b_n ({\bf r}_i)= \frac{1}{\sqrt{N}} \sum_{{\bf k}} \hat{b}_n ({\bf k}) \exp (i{\bf k} {\bf r}_i)
\end{equation}
with the Fourier component
\begin{equation}
\label{fourier}
\hat{b}_n ({\bf k})= \frac{1}{\sqrt{N}} \sum_{{\bf r}_i } b_n ({\bf r}_i) \exp (-i{\bf k} {\bf r}_i)\;,
\end{equation}
where ${\bf k}$ has components $k_j = 2\pi m_j/L, \; m_j = 0,\pm 1, \pm 2,...,\; j=1,2,...,d$ in the range $ - \pi \leq k_j  < \pi  $. Fourier components  follow the relation $\hat{b}_n^*({\bf k})=\hat{b}_n (-{\bf k})$ according to Eq.\ref{fourier}. The correlation function of distance can be calculated as
\begin{equation}
g ({\bf r}, t,L)=\frac{1}{N} \sum_{{\bf r}_i}\sum_{n=1}^{N}b_n ({\bf r}_i) b_n ({\bf r}_i + {\bf r}) \lambda_n\;.
\end{equation}
By using the orthogonal condition
\begin{equation}
\frac{1}{N}\sum_{{\bf r}_i} \exp (i{\bf k}{\bf r}_i) \exp (i{\bf k}'{\bf r}_i)=\delta_{{\bf k}+{\bf k}', 0}
\end{equation}
and the finite-size scaling form of eigenvalues in Eq. (\ref{lambdas}), the correlation function of distance can be expressed as
\begin{equation}
\label{cscaling}
g({\bf r},t,L)=L^{-d+2-\eta}\sum_{n} \sum_{{\bf k}}  f_n ( tL^{1/\nu})\left|\hat{b}_n ({\bf k})\right|^2 \exp (-i {\bf k}{\bf r})\;,
\end{equation}
where ${\bf k}=2\pi {\bf m}/L$. 
 
From this result, we propose that the correlation function of distance follows the finte-size scaling form
\begin{equation}
\label{vscaling}
g({\bf r},t,L)=A\; \left |{\bf r}\right|^{-d+2-\eta} G \left( {\bf r} /L, tL^{1/\nu} \right)\;,
\end{equation}
for $|t| \ll 1$ and $L \gg \tilde{a}$. It is expected that the directional dependence of the universal scaling function $G \left( {\bf r} /L, tL^{1/\nu} \right)$ can be negleted for $|{\bf r}| \ll L$ and becomes unnegligible when $|{\bf r}|$ is comparable to $L$.

From Eq.(\ref{vscaling}), we can get the finite-size scaling of the second moment correlation length
\begin{equation}
\xi (t,L) = \sqrt{\frac{1}{2d}\frac{\sum_{\bf r} \left |{\bf r} \right |^2 g ({\bf r},t,L)}{\sum_{\bf r} g ({\bf r},t,L)}}= L X(tL^{1/\nu})
\end{equation}
in agreement with Eq.(\ref{xi-scaling}). Further, we can obtain the finite-size scaling of susceptibility 
\begin{equation}
\chi (t,L) = L^{\gamma/\nu} f_\chi (t L^{1/\nu})\;,
\end{equation}
with $\gamma/\nu = 2-\eta$.

From correlation functions of different system sizes $L$, we choose the distances $x_\lambda = \lambda L$ along the $x$-direction. The logarithm of correlation function at distance $x_\lambda$ can be written as
\begin{equation}
\label{logc}
\ln g(x_\lambda, t, L) =(-d+2-\eta)\ln L +\ln G(\lambda, tL^{1/\nu})+C\;,
\end{equation}   
where $C=\ln \left(A\lambda^{-d+2-\eta}\right)$. 

At critical point with $t=0$, $\ln g(x_\lambda, 0,L)$ depends on $\ln L$ linearly with slope $-d+2-\eta$. The deviation from this linear dependence of $\ln g(x_\lambda, t, L)$ appears when  $t \neq 0$. We can use this property to fix the critical point and critical exponent $\eta$ of a system.

Using the correlation function at distances $x, 2x, 4x$ along the $x$-direction in the lattice with sizes $L, 2L, 4L$ respectively, we can define a ratio 
\begin{equation}
R=\frac {g(x,t,L) g(4x, t, 4L)}{g(2x,t, 2L)^2}\;.
\end{equation}
This ratio can be written into a scaling form as
\begin{equation}
\label{ratio}
R(\lambda, tL^{1/\nu})=\frac {G(\lambda,tL^{1/\nu}) G(\lambda, 2^{2/\nu}tL^{1/\nu})}{G(\lambda,2^{1/\nu}tL^{1/\nu})^2}\;,
\end{equation}
where $\lambda= x/L$. At critical point $t=0$, we have $R(\lambda, 0)=1$, which is independent of $\lambda$ and $L$. At $t\neq 0$, the ratio $R$ depends on both $\lambda$ and $L$. We can determine the critical point of a system from the fixed point of $R$ as a function of $T$ for different $\lambda$ and $L$.

To verify the finite-size scaling structure of correlation function in Eq.(\ref{vscaling}), we investigate correlation functions of the Ising model and the bond percolation in two-dimensional lattice with PBC using the Monte Carlo simulation. The lattice sizes $L=32$, $64$, and $128$ are taken in our simulations.

\section{Correlation function of Ising model}



At zero external field, the Ising model has the Hamiltonian 
\begin{eqnarray}
\label{hamiltion}
H=-J\sum_{\langle i,j \rangle} S_iS_j,
\end{eqnarray}
where interactions are restricted to the nearest neighbors and spins can take two values, $S_i\ni \{-1,+1\}$.  A configuration of system is characterized by $\{S_i\}=(S_1,S_2,...,S_N)$ with $N=L \times L$.  It has a probability $p(\{S_i\})= e^{- H/k_B T}/Z$ with $Z=\sum_{\{S_i\}} e^{-H/k_B T}$, where the summation is done for all samples of configuration. The Wolff algorithm \cite{Wolff1989} is used to simulate configurations of the Ising model. 

In the average of Eq.\ref{cor}, only configurations with the positive total magnetization are used. If the total magnetization $M$ is negative after a Monte Carlo simulation step, we make a flip $S_i \to - S_i$ to all spins so that $M$ becomes positive agian.  For $N$ spins in the lattice, there are $N(N_1)/2$ correlations between spins. The correlation function $g(\vec{r},t,L)$ is calculated by the average of correlations $C_{ij}$ with $\vec{r}_i - \vec{r}_j =\vec{r}$ as
\begin{equation}
\label{gfun}
g(\vec{r},t,L)=\frac {\sum_{i,j}C_{ij}\delta\left[\vec{r}-(\vec{r}_i-\vec{r}_j)\right]}{\sum_{i,j}\delta\left[\vec{r}-(\vec{r}_i-\vec{r}_j)\right]}
\end{equation}


It has been found that the two-dimensional Ising model in square lattice has the critical point at $k_B T_c/J =2/\ln (1+\sqrt{2})\approx 2.269$\cite{Onsager1944}, and the critical exponents $\nu=1$ and $\eta=\frac 1 4$. To verify the finite-size scaling structure of correlation function, we simulate the Ising model of sizes $L=32, 62,128$ around the critical point with $tL^{1/\nu}= -2, 0, 2$. 

\begin{figure}[th]
\includegraphics[width=8cm,height=4cm]{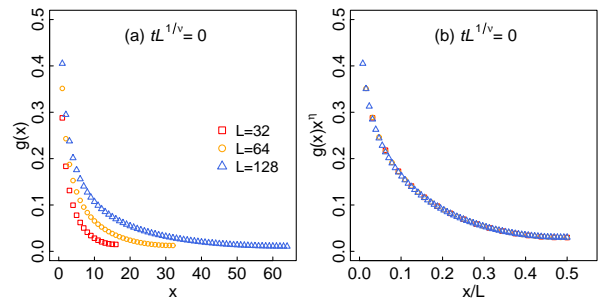}
\caption{\label{fig:ising0} Left: correlation function along $x$-axis of the Ising model at $T_c$. Right: its finite-size scaling function. }
\end{figure}

In Fig.\ref{fig:ising0} (a), the correlation function $g(x,t,L)$ along the $x$-direction is plotted at the critical point and for sizes $L=32, 64, 128$. The scaled correlation function $g(x,t,L) x^\eta$ is shown with respect to the scaled distance $x/L$ in Fig.\ref{fig:ising0} (b), where the curves of different $L$ collapse into one curve. This confirms the finite-size scaling structure of correlation function in Eq. (\ref{vscaling}).

\begin{figure}[th]
\includegraphics[width=8cm,height=4cm]{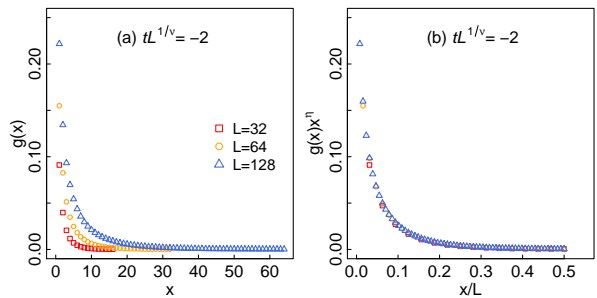}
\caption{\label{fig:isingn} Left: correlation function along $x$-axis of the Ising model below $T_c$. Right: its finite-size scaling function.}
\end{figure}

Below the critical point $T_c$, the Ising model is simulated at sizes $L=32, 64, 128$ and corresponding temperatures with $t L^{1/\nu}=-2$. The results are presented in Fig.\ref{fig:isingn} . The finite-size scaling structure of correlation function is confirmed also at temperatures below $T_c$.

\begin{figure}[th]
\includegraphics[width=8cm,height=4cm]{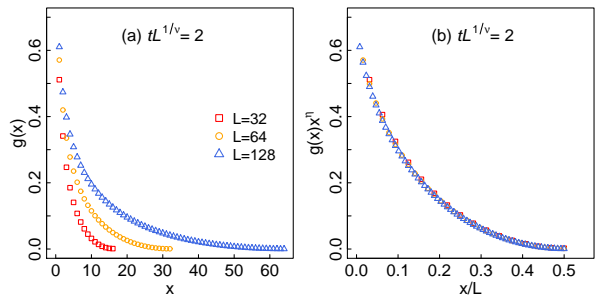}
\caption{\label{fig:isingp}Left: correlation function along $x$-axis of the Ising model above $T_c$. Right: its finite-size scaling function.}
\end{figure}

Above the critical point $T_c$,  the correlation functions of the Ising model are simulated at sizes $L=32, 64, 128$ and corresponding temperatures with $tL^{1/\nu}=2$.  They are shown in Fig.\ref{fig:isingp} and confirm the finite-size scaling structure of correlation function for temperatures above $T_c$.

\begin{figure}[th]
\includegraphics[width=8cm,height=4cm]{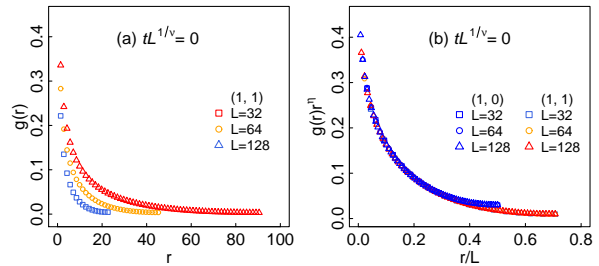}
\caption{\label{fig:xy-ising0} Left: correlation function along diagonal direction of the Ising model at critical point. Right: its finite-size scaling function in comparison with that along $x$-axis}
\end{figure}

\begin{figure}[th]
\includegraphics[width=8cm,height=4cm]{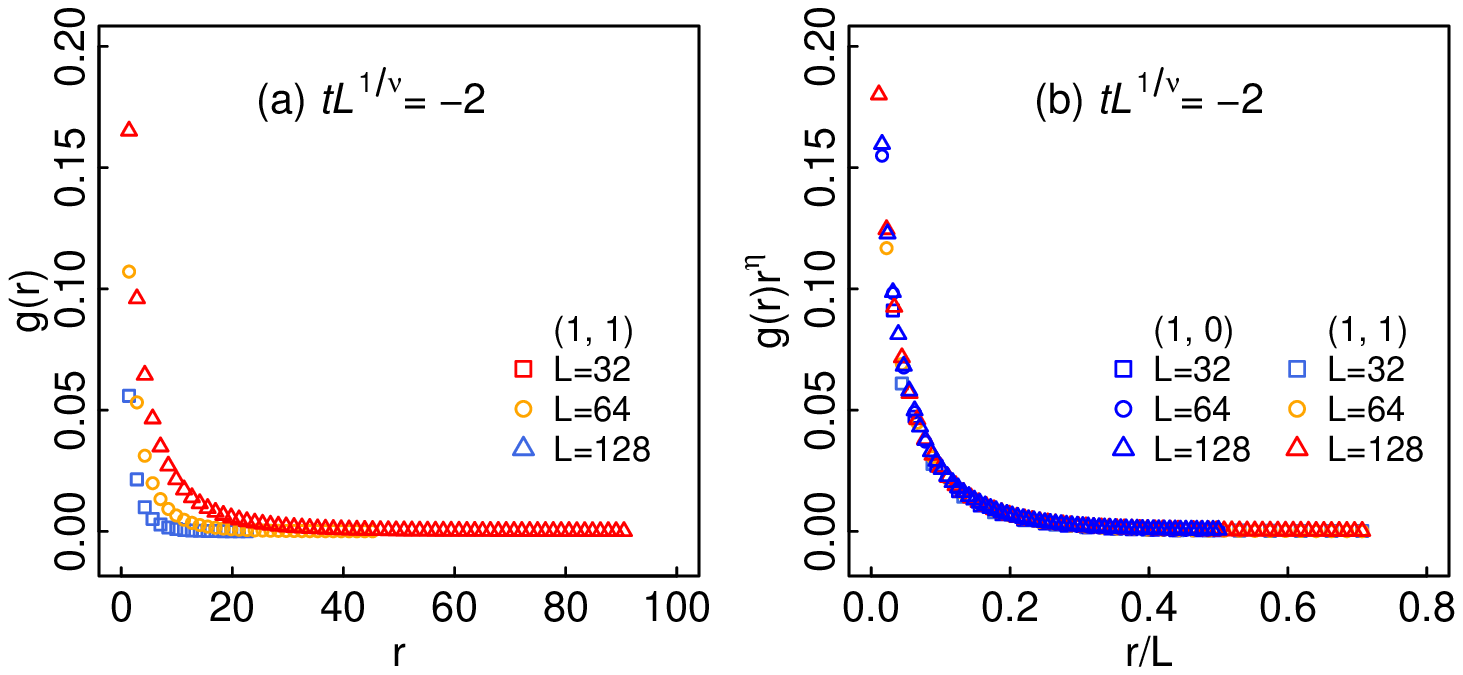}
\caption{\label{fig:xy-ising-2} Left: correlation function along diagonal direction of the Ising model below $T_c$. Right: its finite-size scaling function in comparison with that along $x$-axis.}
\end{figure}

\begin{figure}[th]
\includegraphics[width=8cm,height=4cm]{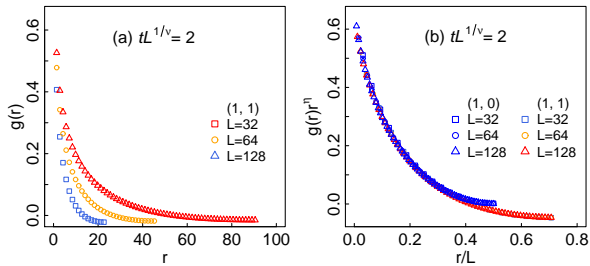}
\caption{\label{fig:xy-ising2} Left: correlation function along diagonal direction of the Ising model above $T_c$. Right: its finite-size scaling function in comparison with that along $x$-axis.}
\end{figure}

At the critical point, the correlation functions along the diagonal direction of lattice with $\vec{r}=  (\frac {r}{\sqrt 2}, \frac {r}{\sqrt 2})$ are presented in Fig.\ref{fig:xy-ising0}. The finite-size scaling structure of correlation function is further confirmed. We can see that the finite-size scaling funtion at distance compariable with $L/2$ becomes dependent on direction and its value in the diagonal direction is different from that in the $x$-direction. Below $T_c$ with $tL^{1/\nu}=-2$, the correltion functions along the diagnal direction are shown on the left and their finite-size scaling function on the right of Fig.\ref{fig:xy-ising-2}. The correlation functions and their finite-size scaling function above $T_c$ with $tL^{1/\nu}=2$ are plotted in Fig.\ref{fig:xy-ising2}. 

\begin{figure*}[h]
\includegraphics[width=8cm,height=12cm]{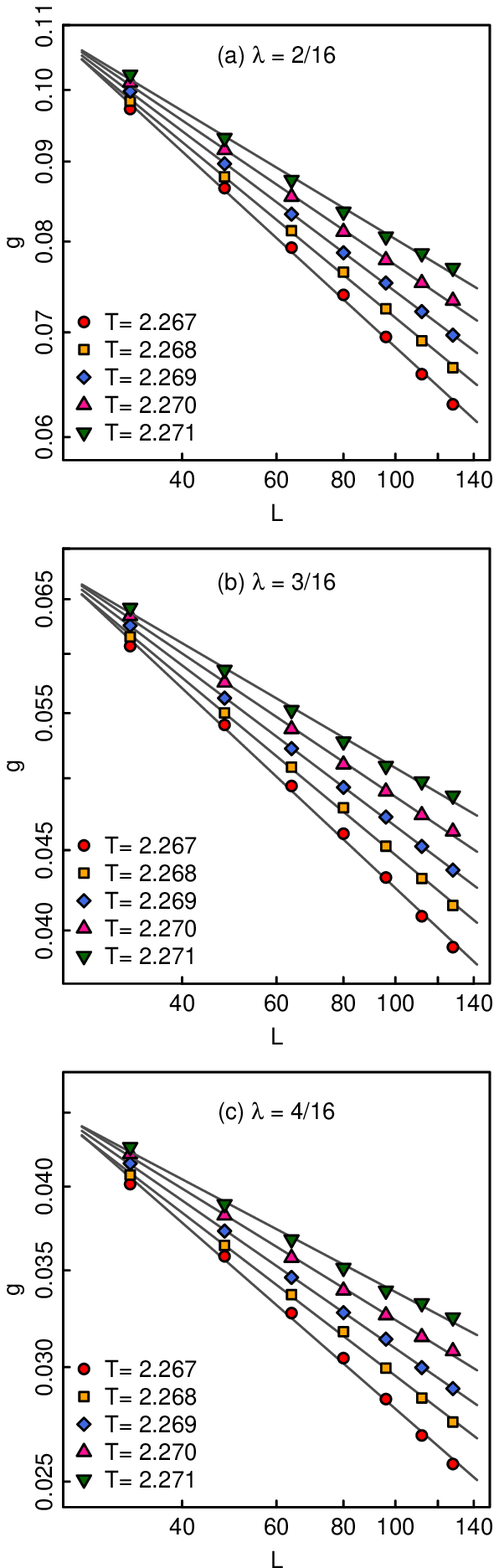}
\caption{\label{fig:ising-tc3} Log-log plot of correlation function of the Ising model along $x$-axis $g(\lambda L, t, L)$ with respect to system size $L$.}
\end{figure*}

The log-log plot of $g(x_\lambda,t,L)$ with respect to $L$ is shown in Fig.\ref{fig:ising-tc3} for $x_\lambda = \lambda L$ with $\lambda = 1/8, 3/16, 1/4$, respectively. At $T < T_c$, the correlation function has nonzero curveture at first. With the increase of temperature, it becomes a straight line at $T_c$ and  is curved again at $T > T_c$.

\begin{figure*}[h]
\includegraphics[width=8cm,height=4cm]{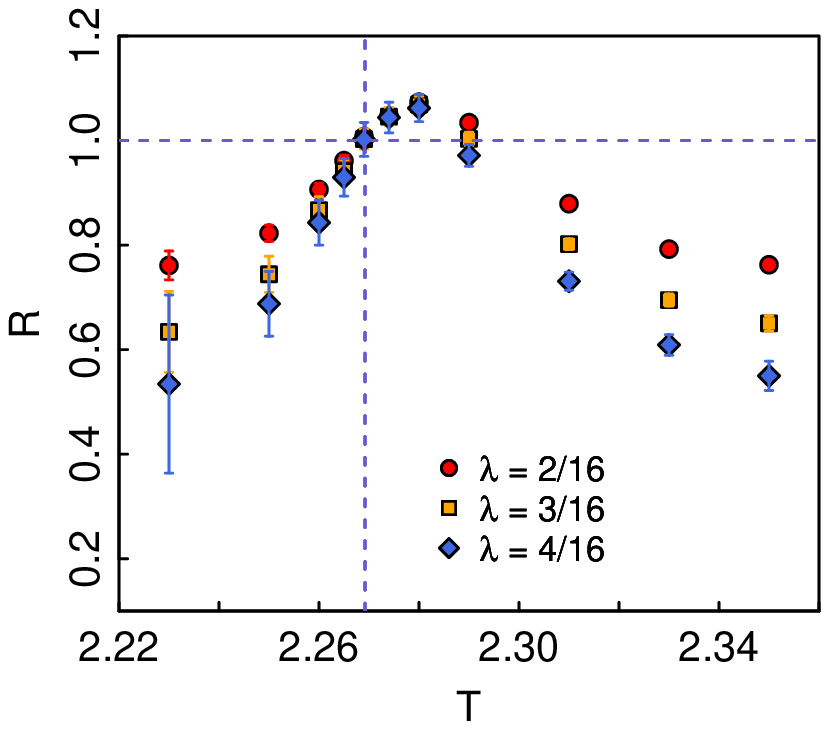}
\caption{\label{fig:ising-tc} Ratio $R(\lambda, tL^{1/\nu})$ of the Ising model with respect to temperature at $L=32$ and different $\lambda$.}
\end{figure*}

In Fig.\ref{fig:ising-tc}, the ratio $R$ is presented as a function of temperature $T$ at $\lambda=1/8, 3/16, 1/4$ and for $L=32$. The different curves of $R$ at different $\lambda$ cross at $T_c$ with $R(\lambda, 0)=1$.

\section{Correlation function of bond percolation} 

In a two-dimensional lattice, bonds are added randomly to connect any two sites in neighbourhood. With the increase of bond number $N_p$, the largest cluster in the lattice becomes larger and larger. When the reduced bond number $p=N_p/N$ reach its critical value $p_c =0.5$, the size of the largest cluster becomes compariable with system size and there is a bond percolation phase transition with critical exponents $\nu=\frac 4 3$ and $\eta=\frac {5} {24}$\cite{Sykes1964}. In one- and two-dimensional lattices, the criticality of networks with long-range connections has been investiated \cite{Yang2016}.

For any configuration in the two-dimensional lattice, two sites $i$ and $j$ are considered to be connected if they belong to the same cluster except for the largest one. The correlation $p_{ij}$ between $i$ and $j$ in a configuration is equal to $1$ when they are connected and otherwise is equal to $0$.
With the average over all configutations, the correlation between sites $i$ and $j$ is calculated as \cite{stauffer1994introduction,Wu1978} 
\begin{equation}
C_{ij} = \langle p_{ij} \rangle \;.
\end{equation}

\begin{figure*}[h]
\includegraphics[width=8cm,height=4cm]{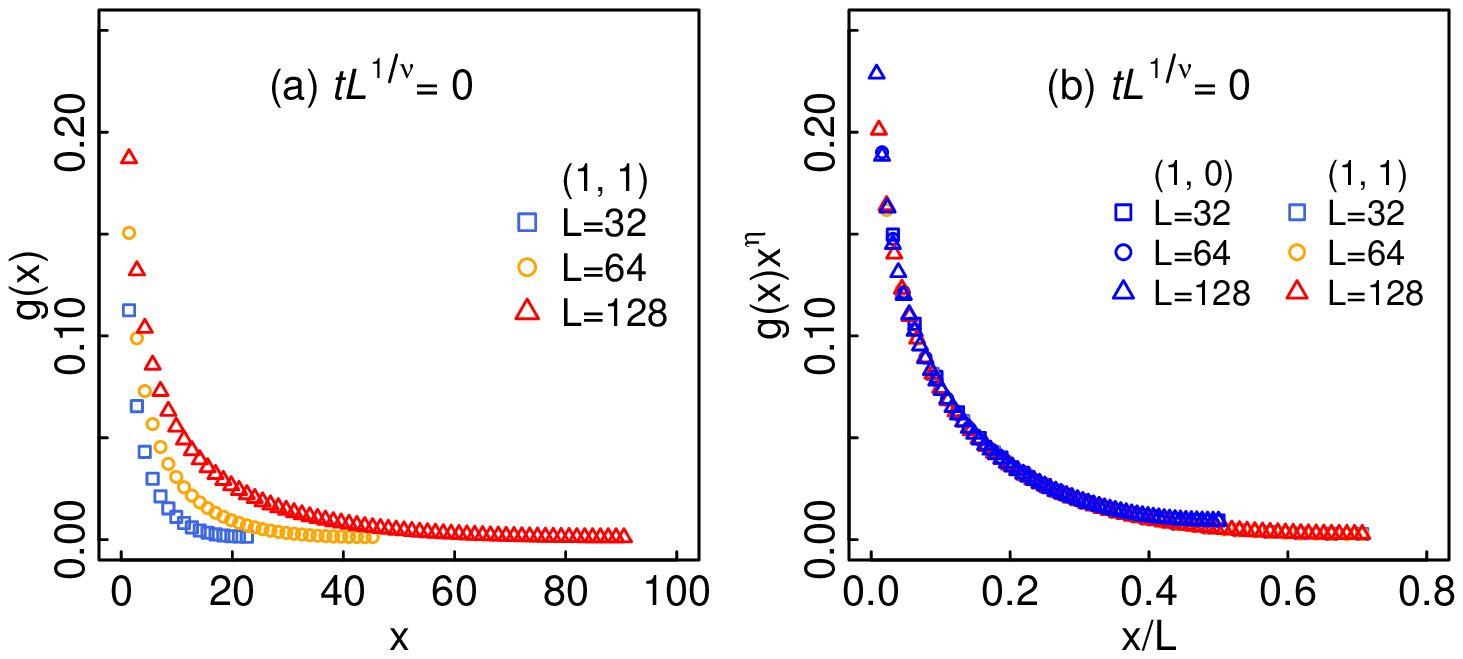}
\caption{\label{fig:xyperc0} Left: correlation function along diagonal direction for bond percolation at $p=p_c$. Right: its finite-size scaling function in comparison with that along $x$-axis.}
\end{figure*}

\begin{figure*}[h]
\includegraphics[width=8cm,height=4cm]{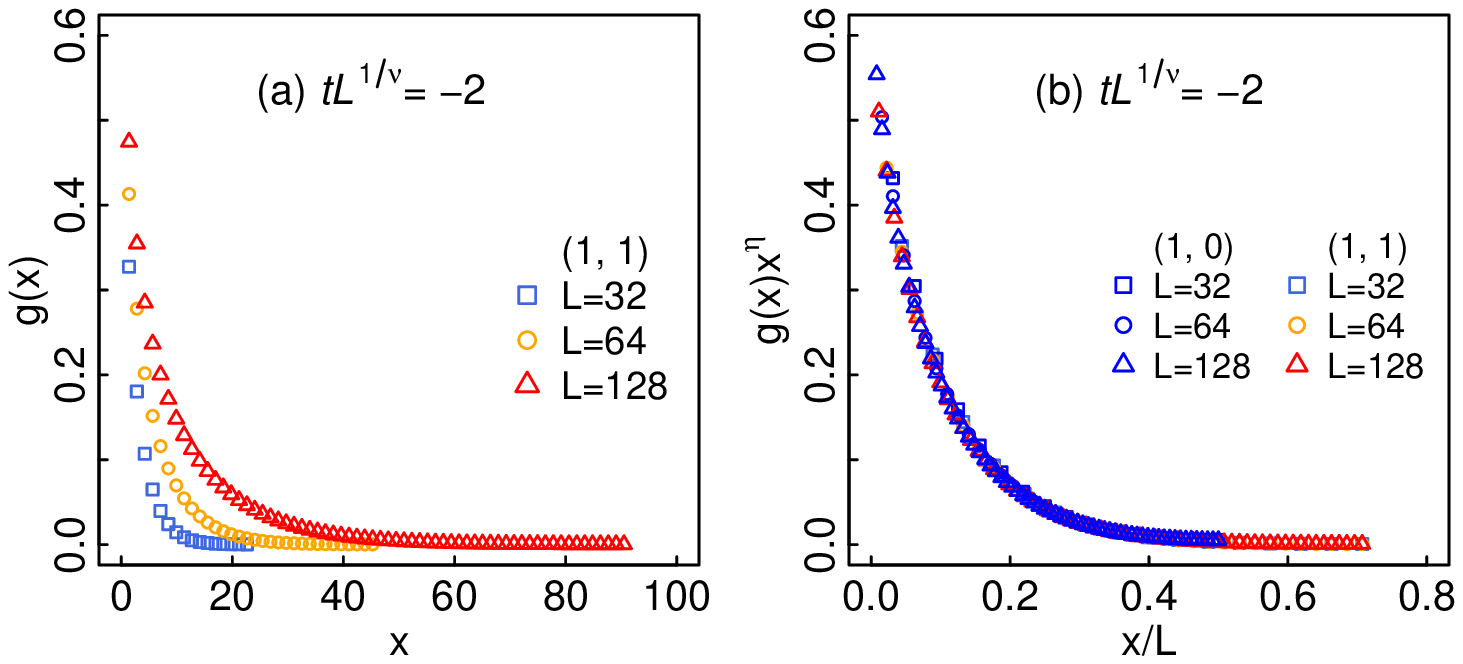}
\caption{\label{fig:xyperc-2} Left: correlation function along diagonal direction for bond percolation at $p < p_c$. Right: its finite-size scaling function in comparison with that along $x$-axis.}
\end{figure*}

\begin{figure*}[h]
\includegraphics[width=8cm,height=4cm]{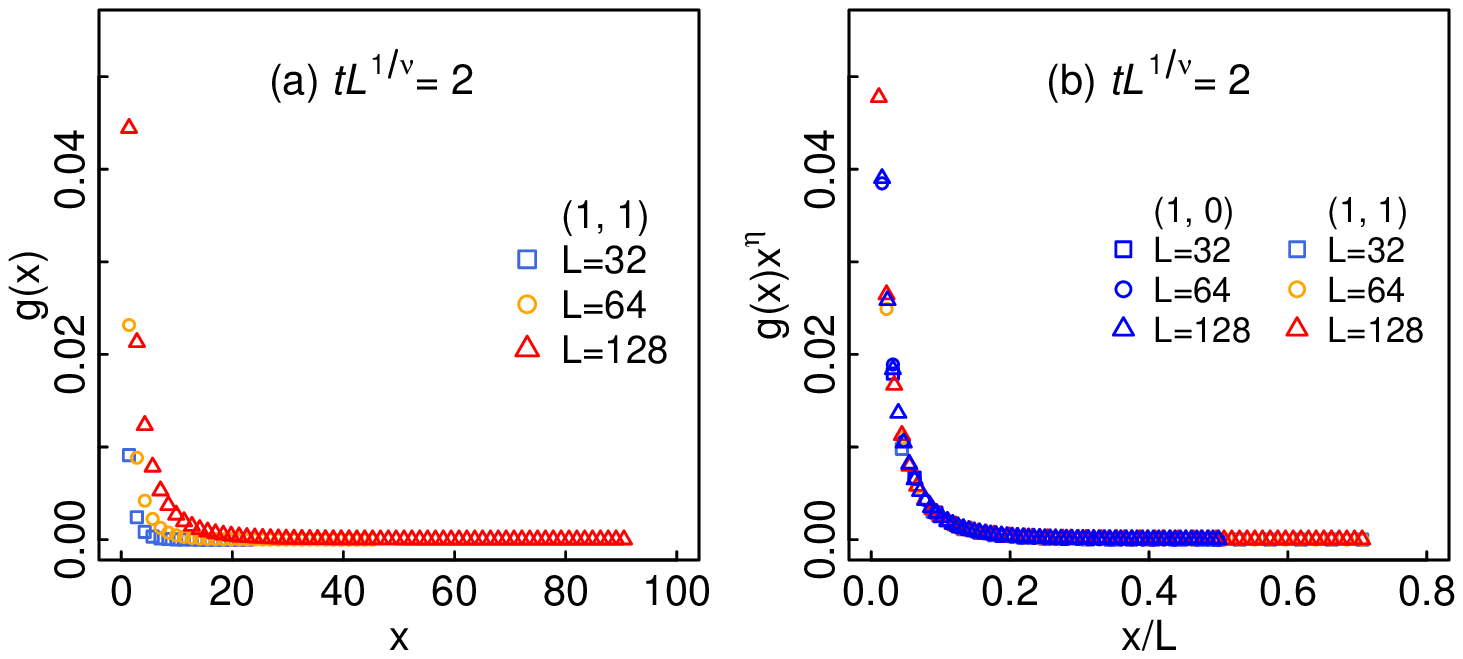}
\caption{\label{fig:xyperc2} Left: correlation function along diagonal direction of bond percolation at $p > p_c$. Right: its finite-size scaling function in comparison with that along $x$-axis.}
\end{figure*}

Using the defnition in Eq.\ref{gfun}, we can calculate the correlation function $g(\vec{r},t,L)$ of two-dimensional bond percolation from $C_{ij}$. The results at $T_c$ are shown in Fig.\ref{fig:xyperc0}. At the left side, the correlation functions along diagonal direction of different $L$ are shown. Their finite-size scaling functions are presented at the right side and are compared with that along $x$-direction. Furhter, the correlation functions of bond percolation below and above $p_c$ are demonstrated in Figs.\ref{fig:xyperc-2} and \ref{fig:xyperc2}, respectively. The finite-size struction of correlation function in Eq.\ref{vscaling} has been also confirmed by the correlation function of bond percolation. 

\begin{figure*}[h]
\includegraphics[width=8cm,height=12cm]{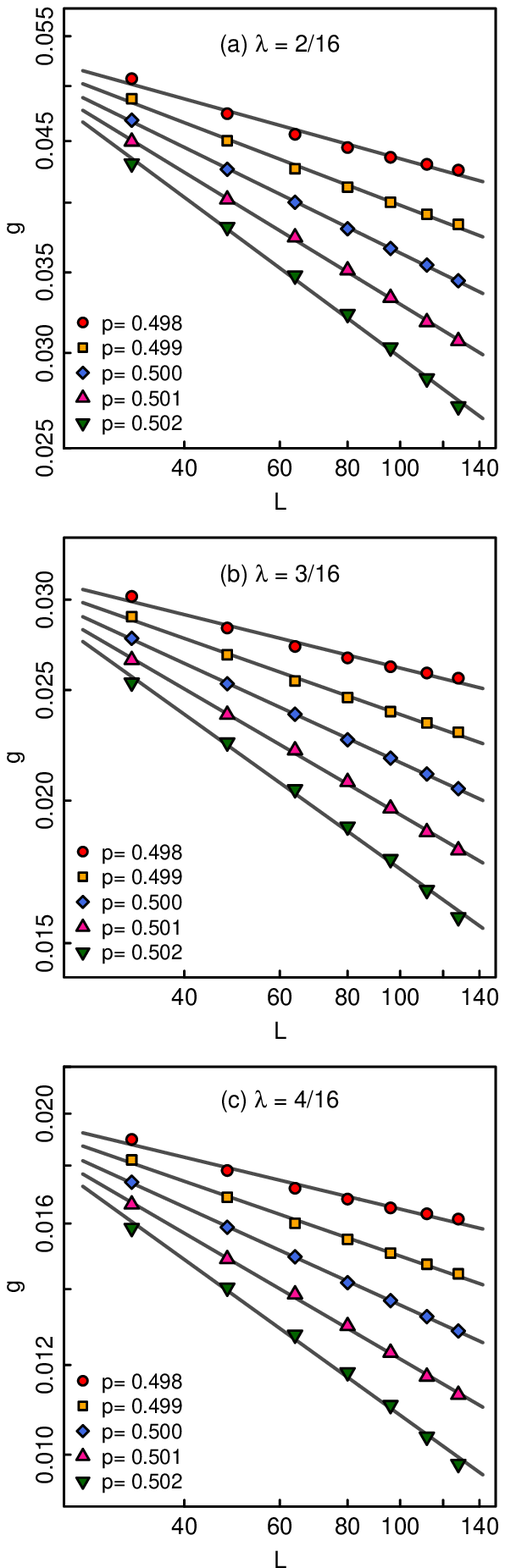}
\caption{\label{fig:perc-tc3} Log-log plot of correlation function along $x$-axis for bond percolation $g(\lambda L, t, L)$ with respect to system size $L$.} 
\end{figure*}

\begin{figure*}[h]
\includegraphics[width=8cm,height=4cm]{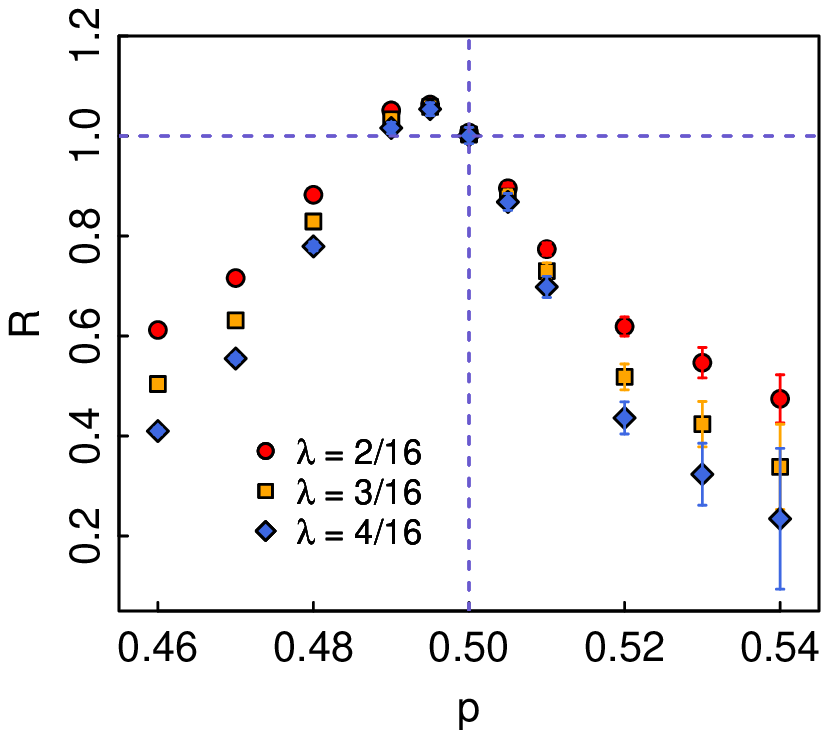}
\caption{\label{fig:perc-tc} Ratio $R(\lambda, tL^{1/\nu})$ of bond percolation with respect to $p$ at $L=32$ and different $\lambda$.}
\end{figure*}

To vertify the finite-size structure in Eq.\ref{logc}, the log-log plot of $g(x_\lambda, t, L)$ with respect to $L$ is shown in Fig.\ref{fig:perc-tc3} with $t=(p-p_c)/p_c$. With $\lambda=1/8, 3/16, 1/4$ respectively,  the correlation functions at $x_\lambda =\lambda L$ are plotted for $p=0.498, 0.499, 0.500, 0.501, 0.502$. With the increase of $p$, $\ln g(x_\lambda, t, L)$ as a function of $\ln L$ is curved downward at the beginning and becomes curved upward finally. It is at $p_c=0.500$ that $\ln g(x_\lambda, t, L)$ depends on $\ln L$ linearly. In Fig.\ref{fig:perc-tc}, the ratio $R$ defined in Eq.\ref{ratio} is shown with respect to $p$ for $\lambda=1/8, 3/16, 1/4$. It is found that $R$ is independent of $\lambda$ and equal to $1$ at $p_c$. Therefore, the finite-size scaling of correlation function has been confirmed by the bond percolation in two-dimensional lattice.

\section{Conclusions}

We propose here the finite-size scaling of correlation function in a finite system near its critical point.  For the finite system with size $L$ and at the reduced temperature $t=(T-T_c)/T_c$, its correlation function at distnce ${\bf r}$ can be scaled as $g({\bf r},t,L) = A|{\bf r}|^{-(d-2+\eta)} G({\bf r}/L, tL^{1/\nu})$. At the distance $|{\bf r}| \ll L$, the finite-size scaling function $G({\bf r}/L, tL^{1/\nu})$ is independent of the direction of ${\bf r}$. When the distance becomes compariable with $L$, the directional dependence of $G({\bf r}/L, tL^{1/\nu})$ is nonnegligible. From the finite-size scaling of correlation function, we can obtain the finite-scalings of susceptibility \cite{Privman1991} and of the second moment correlation length \cite{{PhysRevB.30.322}}.

Using Monte Carlo simulation, the correlation functions of two-dimensional Ising model and bond percolation in square lattices are calculated. These results of both Ising model and bond percolation verify the finite-size scaling of correlation function proposed above.

We thank the financial support by Key Research Program of Frontier Sciences, CAS (Grant No. QYZD-SSW-SYS019). Yongwen Zhang thanks the postdoctoral fellowship funded by the Kunming University of Science and Technology. 

\bibliography{reference}

\end{document}